\documentclass{article}

\usepackage{arxiv}

\usepackage[utf8]{inputenc} 
\usepackage[T1]{fontenc}    
\usepackage{hyperref}       
\usepackage{url}            
\usepackage{booktabs}       
\usepackage{amsfonts}       
\usepackage{nicefrac}       
\usepackage{microtype}      
\usepackage{lipsum}
\usepackage{color}
\usepackage{algorithm}
\usepackage{algorithmic}
\usepackage{amsmath,amsfonts,amssymb}
\usepackage{subfigure}
\usepackage{lineno}
\usepackage{graphicx}
\usepackage{bm}
\graphicspath{{Figures/}}
\usepackage{subfigure}
\usepackage{newtxtext}
\usepackage{newtxmath}
\newcommand{\tn}[1]{\ensuremath{\underline{\mathsfbi{#1}}}}
\DeclareMathAlphabet{\mathsfbi}{OML}{ntxsfmi}{b}{it}
\usepackage[OMLmathsans]{isomath}

\title{Data-driven time-dependent bases for turbulent airfoil wake-extreme gust interactions}

\renewcommand{\shorttitle}{Zamani Ashtiani \& Fukami / Data-driven time-dependent bases for extreme aerodynamics}

\author{
Shaghayegh Zamani Ashtiani$^{[1,2]}$ and Kai Fukami$^{[1,*]}$
\\
\\
1. Department of Aerospace Engineering, Graduate School of Engineering, Tohoku University, Sendai, 980-8579, Japan\\
2. Department of Mechanical Engineering and Materials Science, University of Pittsburgh, PA 15260, USA\\
\\
$^*$ Corresponding author: \texttt{kfukami1@tohoku.ac.jp}
}

\begin{document}
\maketitle

\begin{abstract}
We analyze interactions between turbulent airfoil wakes and extreme gusts using a data-driven framework with time-dependent bases.
The current approach represents each snapshot with time-varying bases consisting of two-dimensional in-plane modes and one-dimensional spanwise modes, together with a reduced covariance matrix. 
By deriving closed-form evolution equations, we advance these components using a small rolling window, eliminating the need for full-history storage.
Applied to extreme vortex gust-airfoil interactions at $Re=5000$, we reveal that the first in-plane mode dominates before impingement, while the second mode gains energy post-impingement, amplified by gust intensity.
During impingement, leading modes exhibit multiscale features such as shear-layer roll-up and fine-scale turbulence.
The temporal evolution of the mode-energy spectrum and its rank gaps further quantifies these interactions. 
A large leading-mode gap indicates energy concentrated in a few coherent structures, leading to faster recovery, whereas a smaller gap indicates energy distributed across many modes, consistent with richer multiscale activity and delayed re-stabilization.
These trends also follow the transient lift dynamics, with higher amplitude and more oscillations indicated by a rise in the leading singular values. 
This work may provide a foundation for time-varying data-driven modal analysis of extreme gust encounters.

\end{abstract}

\section{Introduction} 
\label{sec:intro}

Understanding gust--wing interaction is crucial for flight safety and design. { As the gust intensity $G$ increases, the aerodynamics of these interactions become increasingly extreme. The gust intensity is defined as $G \equiv u_g/u_\infty$, where $u_g$ is the characteristic gust velocity and $u_\infty$ is the free-stream speed.  For lower gust ratios $(G < 1)$, a substantial body of literature exists, including both experimental and computational studies \cite{rockwell1998vortex, fukami2024data}. However,} small aircraft frequently encounter high gust ratios $(G \ge 1)$ in urban canyons, mountainous terrain, and severe turbulence \cite{jones2022physics,fukami2023grasping}.
In such a violent airspace, separation and the wake structure are reorganized, driving large, rapid changes. Particularly when such a strong gust interacts with {the} turbulent wake behind a wing, the flow transitions from a {statistically steady} baseline to a strongly nonlinear, {transient} response, marked by three-dimensional vortex breakdown, spanwise instabilities, fine-scale turbulence, and sharply fluctuating aerodynamic forces, which makes physical interpretation and analysis challenging \cite{fukami2025extreme,Taira_XAero_review}. 
{This complexity motivates compact low-rank representations that remain interpretable by tracking the dominant flow motions and the evolving distribution of modal energy over time.}

To make such complex aerodynamic flows analyzable, one can consider data-driven reduced-order models that project high-dimensional flow fields onto low-dimensional subspaces. 
Among them, proper orthogonal decomposition (POD)~\cite{lumley1967structure,sirovich1987turbulence} identifies an energy-optimal set of orthogonal modes that capture most of the variance from snapshots, and dynamic mode decomposition (DMD)~\cite{schmid2010dynamic} extracts spatiotemporal modes with single frequencies and growth/decay rates that approximate linear evolution between snapshots. 
These approaches perform well for statistically stationary unsteady flows. 
However, performance with their standard forms degrades in highly transient regimes (e.g. strong gusts) because POD relies on fixed spatial bases, and DMD assumes one time-invariant linear map over the analysis window \cite{linot2025extracting}.

{To preserve transient dynamics, space-time POD has been used, which defines orthogonal modes jointly over space and a finite time window, enabling each mode to capture an evolving spatiotemporal pattern \cite{lumley1967structure,lumey2012stochastic}. 
In the limit of a very short window, space-time POD approaches the standard snapshot (space-only) POD \cite{frame2023space}. 
Conditional space-time POD further refines the approach by conditioning the analysis while selecting space-time windows according to an external event or parameter \cite{schmidt2019conditional}. 
The performance of space-time POD depends on the choice of window length and the degree of statistical convergence attainable with the available data.
Spectral POD can also be used to model statistically steady flows in the frequency domain \cite{lumley1967structure,lumey2012stochastic}.  
At each frequency, spectral POD yields an energy-optimal set of spatial modes that represents the flow content at that frequency more accurately, on average, than any other basis of the same rank. 
Spectral POD can be obtained from a space-time POD formulation in the limit of a long time window for statistically stationary flows \cite{towne2018spectral,frame2023space}.} {In addition to POD-based space--time formulations, multi-resolution DMD has been applied to a range of unsteady flows, yielding modes that are localized in time and separated by characteristic time scales \cite{kutz2016multiresolution}.}

Particularly for extreme gust impingement on airfoils, recent studies have revealed that nonlinear autoencoders enable the extraction of low-dimensional manifolds that capture high-dimensional vortical flow physics while associating with aerodynamic forces in a reduced-order manner \cite{fukami2023grasping,fukami2024data}. 
{From an information-theoretic viewpoint, the encoder can be interpreted as a compression map that aims to reconstruct modal structure associated with future aerodynamic response  while discarding redundant variability \cite{fukami2025information,koshikawa2026convolutional}}. However, nonlinear machine learning can generally require substantial training data, offline training time, and careful architectural choices based on prior knowledge of flow physics.

{Recent reduced-order model techniques obtain time-varying structure, either from the governing equations (model-driven) or directly from data (data-driven). In the model-driven formulations, the solution is constrained to a time-dependent low-dimensional subspace and the governing equations are projected to derive coupled evolution equations for the bases and modal coefficients. Dynamically orthogonal decomposition, biorthogonal decomposition, and dynamically biorthogonal decompositions are examples of model-driven reduced-order models that have been applied to high-dimensional stochastic partial differential equations \cite{sapsis2009dynamically,cheng2013dynamically,patil2020real} and other application areas, including combustion, linear sensitivity analysis, and dynamical instability studies \cite{RNB21,babaee2016minimization,donello2022computing,amiri2024time}. 
These decompositions are closely related, but differ in their constraints and formulations; in practice, numerical implementation choices can affect their performance. In the data-driven approach, the time-dependent bases evolution is inferred from instantaneous time derivatives of streaming data, thereby avoiding full-history storage while preserving spatiotemporal features and enabling \emph{in situ} compression of multidimensional fields \cite{ashtiani2022scalable}. In addition to data compression, the explicit tracking of the evolving low-rank structure in time is particularly attractive for modal analysis in complex phenomena such as extreme vortex-gust airfoil interaction. In this study, we tailor this data-driven, time-dependent bases framework to extreme vortex--gust airfoil interactions to extract time-varying low-rank structure.}

{This study aims to reveal the transient dynamics of extreme vortex--gust encounters through time-varying modal analysis with an unsteady base state, while providing a low-dimensional characterization of the complex vortex gust--airfoil interaction. To this end,} we consider large-eddy simulations of a NACA0012 airfoil at $\alpha=14^\circ$ and a chord-based Reynolds number of 5000 with an extreme vortex gust encounter of \( G  \ge 1\). 
The current study reveals time-varying dominant characteristics of the extreme aerodynamics under turbulent wake conditions as transient modes and singular-value dynamics. 
The current paper is organized as follows: the method is described in section \ref{sec:method}; results are presented in section \ref{sec:results}; and conclusions are given in section \ref{sec:conclusion}.

\section{Method} \label{sec:method}
To reveal the time-varying dominant flow physics of extreme vortex--gust airfoil interactions, we use a time-dependent, low-rank representation of the flow field. 
The field is decomposed into time-varying in-plane and spanwise modes whose modal energies quantify their relative contributions. 
This reduced representation provides an interpretation of the governing physics and flow organization in a time-varying manner. We represent the vortex–gust airfoil interaction field $q(x,y,z,t)$ by its low-rank approximation,
\begin{equation}
q(x,y,z,t)=\hat{q}(x,y,z,t)+\epsilon(x,y,z,t),
\label{eq1}
\end{equation}
where $\hat{q}(x,y,z,t)$ is the low-rank representation of the field and $\epsilon(x,y,z,t)$ is the truncation error. 
This residual denotes the portion of the true field not represented by the low-rank approximation and thus quantifies the unresolved flow energy.

The time-dependent bases framework supports multidimensional data and multiple decomposition techniques. Based on the vortex–gust interaction physics, two approaches can be considered.
One fully decomposes the field into one-dimensional modes in the \(x\), \(y\), and \(z\) directions, offering a high compression ratio and interpretation via modal energy. 
{In this scenario, the time-dependent bases take the following form,
 \begin{equation}\label{eq:mother3D}
 \hat{q}(x,y,z,t) =  \sum_{{i_3}=1}^{r_3} \sum_{{i_2}=1}^{r_2} \sum_{{i_1}=1}^{r_1} 
 \tn{\mathsfbi{T}}_{i_1 i_2 i_3}(t)\xi_{i_1}(x,t)\xi_{i_2}(y,t) \xi_{i_3}(z,t).
\end{equation}
Where $\xi_{i_1}(x,t)$, $\xi_{i_2}(y,t)$, and $\xi_{i_3}(z,t)$ are one-dimensional modes in the $x$, $y$, and $z$ directions, respectively, and $\tn{\mathsfbi{T}}_{i_1 i_2 i_3}(t)$ is the core tensor. 
In this formulation, the one-dimensional modes are not directly interpretable as flow structures; physical features are revealed only after reconstructing the full field. }
Further details on this approach are referred to Zamani Ashtiani et al.~\cite{ashtiani2022scalable}.

\begin{figure}[t]
\includegraphics[width=\linewidth]{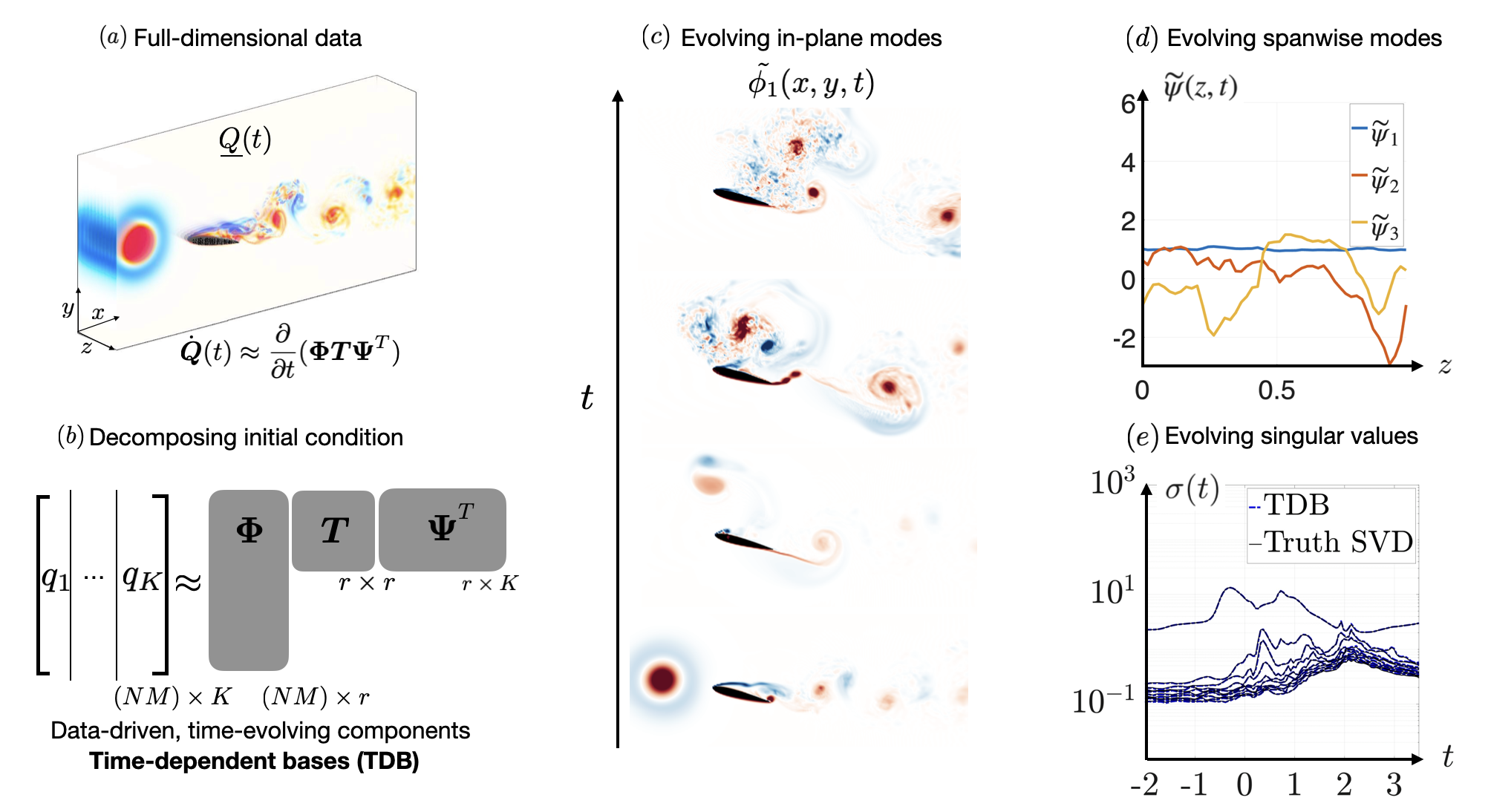}
\caption{Time-dependent bases workflow: 
$(a)$ snapshot field, $(b)$ matricization and SVD initialization, $(c)$ in-plane modes, $(d)$ spanwise modes, and $(e)$ singular values.}
\label{fig:TDB}
\end{figure}

In response, this study considers adopting the second approach --- the \((x,y)\) plane is grouped into two-dimensional in-plane modes and use one-dimensional spanwise modes in the \(z\) direction.
This typically yields lower compression because the physical space $(x,y)$ is not decomposed. However, it exhibits slower error growth, and the in-plane $(x,y)$ modes explicitly reveal flow structure, including vortices, shear layers, and separation.
The latter is particularly advantageous for the current modal analysis of extreme vortex-airfoil interactions because it aligns the bases with dominant in-plane vortex dynamics, capturing flow structures in the \((x,y)\) plane while representing spanwise variability compactly along the spanwise direction. Consequently, this decomposition reveals the dominant flow physics through in-plane modes and their modal energies, enabling the analysis of the present turbulent airfoil wake-extreme vortex gust interactions, as summarized in figure~\ref{fig:TDB}.
The low-rank approximation is described as,
\begin{equation} 
  \hat{q}(x,y,z,t) =  \sum_{j=1}^{r} \sum_{i=1}^{r} \mathsfbi{T}_{i j}(t)\phi_{i}(x,y,t)\psi_{j}(z,t),  \label{eq:TDB}
 \end{equation}
{with} \(\phi_i(x,y,t)\) are in-plane orthonormal modes, \(\psi_i(z,t)\) are spanwise orthonormal modes, \(\mathsfbi{T}(t) {\in \mathcal{R}^{r \times r}}\) is factorization of the reduced covariance matrix, i.e. \(\mathsfbi{C}(t) = \mathsfbi{T}(t) \mathsfbi{T}^{\mathrm{T}}(t)\), where \(\mathsfbi{C}(t) \in \mathcal{R}^{r \times r} \), and \(r\) denotes the truncation rank. The discrete form of the field is represented on an \(N\times M\times K\) grid, yielding the tensor \(\tn{Q}(t)\in\mathcal{R}^{N\times M\times K}\), where \(N,M,K\) are the numbers of grid points in \(x\), \(y\), and \(z\), respectively. 
We matricize the three-dimensional field into $\mathsfbi{Q}(t)\in\mathcal{R}^{(NM)\times K}$ so that columns encode in-plane $(x,y)$ coherent structures and rows encode spanwise $(z)$ organization, as illustrated in figures \ref{fig:TDB}$(a)$ and $(b)$. 
For simplicity, we use the discrete notation,
\begin{align}
  \hat{\mathsfbi{Q}}(t) &=\mathsfbi{\Phi}(t)\,\mathsfbi{T}(t)\,\mathsfbi{\Psi}^{\mathrm{T}}(t), \quad  \hat{\mathsfbi{Q}}(t)  \in \mathcal{R}^{(NM) \times K}, \quad \mbox{where} \nonumber\\ 
  \mathsfbi{\Phi}(t) &= \big[\, \boldsymbol{\phi}_1(t)\ \big|\ \boldsymbol{\phi}_2(t)\ \big|\ \cdots\ \big|\ \boldsymbol{\phi}_r(t)\,\big], \quad \mathsfbi{\Phi}(t)
  \in \mathcal{R}^{(N M)\times r}, \nonumber\\ 
  \mathsfbi{\Psi}(t) &= \big[\,\boldsymbol{\psi}_1(t)\ \big|\ \boldsymbol{\psi}_2(t)\ \big|\ \cdots\ \big|\ \boldsymbol{\psi}_r(t)\,\big], \quad
\mathsfbi{\Psi}(t) \in \mathcal{R}^{K\times r}.\label{Psi}\
\end{align}
The objective of the time-dependent bases framework is to compute the decomposed components as functions of time. To this end, we derive evolution equations using time derivatives, evaluated via finite differences over a sliding window of snapshots, enabling out-of-core processing without retaining the full time history in memory.

To derive the time-dependent bases equations, we define the inner products between two fields in the in-plane (\(x,y\)) and spanwise (\(z\)) spaces as 
\begin{align}
\langle a,b\rangle_{xy}
&= \iint_{D_{xy}} a(x,y)b(x,y)\,dx\,dy 
\approx \sum_{j=1}^{M}\sum_{i=1}^{N}\! w_{x,i}w_{y,j}\,a(x_i,y_j)b(x_i,y_j)
= \boldsymbol{a}^{\mathrm T}\mathsfbi{W}_{xy}\,\boldsymbol{b},\nonumber\\
\langle c,g\rangle_{z}
&= \int_{D_z} c(z)g(z)\,dz
\approx \sum_{i=1}^{K} w_{z,i}\,c(z_i)\,g(z_i)
= \boldsymbol{c}^{\mathrm T}\mathsfbi{W}_{z}\,\boldsymbol{g},
\end{align}
{in which} $\boldsymbol{w}_x=\big[w_{x_1},\ldots,w_{x_N}\big]$, 
$\boldsymbol{w}_y=\big[w_{y_1},\ldots,w_{y_M}\big]$, and 
$\boldsymbol{w}_z=\big[w_{z_1},\ldots,w_{z_K}\big]$ are the quadrature-weight vectors in the $x$, $y$, and $z$ directions, respectively. Accordingly,
$\mathsfbi{W}_{xy}=\mathrm{diag}~\!\big(\boldsymbol{w}_y\otimes\boldsymbol{w}_x\big)$
and
$\mathsfbi{W}_{z}=\mathrm{diag}(\boldsymbol{w}_z)$. With the inner products defined above, the columns of $\mathsfbi{\Phi}(t)$ and $\mathsfbi{\Psi}(t)$ are sets of orthonormal spatial modes such that
\begin{align}
\mathsfbi{\Phi}(t)^{\mathrm T} \mathsfbi{W}_{xy} \mathsfbi{\Phi}(t) =\mathsfbi{I}, 
~~~\mathsfbi{\Psi}(t)^{\mathrm T} \mathsfbi{W}_{z} \mathsfbi{\Psi}(t) =\mathsfbi{I}. \label{PsiOr}
\end{align}
{In addition, following the time-dependent bases formulation \cite{patil2020real,sapsis2009dynamically}, the time derivatives $\dot{\mathsfbi{\Phi}}(t)$ and $\dot{\mathsfbi{\Psi}}(t)$ are chosen to be orthogonal to the spatial subspaces spanned by the columns of $\mathsfbi{\Phi}(t)$ and $\mathsfbi{\Psi}(t)$, respectively (dynamical orthogonality constraint) to derive a closed-form evolution equation for the decomposed components.
In other words, the time derivatives of these modes are orthogonal to the subspaces they span such that}
\begin{align}
\mathsfbi{\Phi}(t)^{\mathrm T} \mathsfbi{W}_{xy} \dot{\mathsfbi{\Phi}}(t)  =\mathsfbi{0}, 
~~~\mathsfbi{\Psi}(t)^{\mathrm T} \mathsfbi{W}_{z} \dot{\mathsfbi{\Psi}}(t)  =\mathsfbi{0}.\label{Psi0}
\end{align}

{To drive closed-form evolution equations for the decomposed components ($\mathsfbi{\Phi}, \mathsfbi{T}, \mathsfbi{\Psi}$), we take the time-derivative of time-dependent bases decomposition,
$
\dot{\mathsfbi{Q}} = \dot{\mathsfbi{\Phi}} \mathsfbi{T} \mathsfbi{\Psi}^{{\mathrm{T}}} + \mathsfbi{\Phi} \dot{\mathsfbi{T}} \mathsfbi{\Psi}^{{\mathrm{T}}} + \mathsfbi{\Phi} \mathsfbi{T} \dot{\mathsfbi{\Psi}}^{{\mathrm{T}}}
$.
The evolution equations are obtained by taking inner products between $\dot{\mathsfbi{Q}}$ and the in-plane/spanwise bases.
The orthonormality and dynamical orthogonality constraints are then enforced on the inner products~\cite{ashtiani2022scalable}.
In particular, $\langle \mathsfbi{\Psi},\dot{\mathsfbi{Q}}^{\mathrm T}\rangle_{z}$ yields the evolution equation for the in-plane bases $\mathsfbi{\Phi}$, while $\langle \mathsfbi{\Phi},\dot{\mathsfbi{Q}}\rangle_{xy}$ produces the evolution equation for the spanwise bases $\mathsfbi{\Psi}$. 
The evolution equation for the reduced covariance matrix $\mathsfbi{T}$ follows analogously by projecting $\dot{\mathsfbi{Q}}$ onto both $\mathsfbi{\Phi}$ and $\mathsfbi{\Psi}$ under the same constraints. Therefore, these inner-product relations and constraints lead to a closed set of evolution equations for the time-dependent bases, given by the following.}
\begin{align}
 \dot{\mathsfbi{\Phi}} &=\left(\mathsfbi{I}-\mathsfbi{\Phi \Phi}^\mathrm{T} \mathsfbi{W}_{xy}\right) \dot{\mathsfbi{Q}}\mathsfbi{ W}_{z} \mathsfbi{\Psi} \boldsymbol{T}^{-1}, \label{Phi_TDB} \\
~~~ \dot{\mathsfbi{\Psi}} &=\left(\mathsfbi{I}-\mathsfbi{\Psi \Psi}^\mathrm{T} \mathsfbi{W}_{z}\right) \dot{\mathsfbi{Q}}^\mathrm{T} \mathsfbi{W}_{xy} \mathsfbi{\Phi} \boldsymbol{T}^{-{\mathrm{T}}}, \label{Psi_TDB}\\
 \dot{\mathsfbi{T}} &=\mathsfbi{\Phi}^\mathrm{T} \mathsfbi{W}_{xy} \dot{\mathsfbi{Q}} \mathsfbi{W}_{z} \mathsfbi{\Psi}.
~~~~~~~~~~~~~~~~~~~~~~~~~~~~~~~~~~~~~~~~
\label{T_TDB} 
\end{align}
The equations above are equivalent to the dynamically bi-orthonormal decomposition \cite{patil2020real,RNB21,naderi2023adaptive} formulated in a model-driven based framework. 
To integrate the equations in time, we use the fourth-order Runge–Kutta scheme.

The reduced covariance matrix (equation~\ref{T_TDB}) quantifies the importance of dominant flow structures via the singular values \(\sigma_i(t)\). The fraction of energy captured by the first \(r\) modes is \(\sum_{i=1}^{r}\sigma_i(t)^{2}\big/\sum_{i}\sigma_i(t)^{2}\), accordingly growth or decay of \(\sigma_i(t)\) tracks modal energy transfer. 
To extract these values, we perform the singular value decomposition (SVD) of \(\mathsfbi{T}\) as \(\mathsfbi{T}=\mathsfbi{U}\,\mathsfbi{\Sigma}\,\mathsfbi{Y}^{\mathsf{T}}\), where \(\mathsfbi{\Sigma}=\mathrm{diag}(\sigma_1,\dots,\sigma_r)\) with \(\sigma_1 > \sigma_2 > \cdots > \sigma_r\), and \(\mathsfbi{U}\) and \(\mathsfbi{Y}\) contain the left and right singular vectors, respectively. {The time-dependent bases framework is closely related to the instantaneous singular value decomposition of the full data matrix $\mathsfbi{Q}(t)$;
therefore, the diagonal entries of $\mathsfbi{\Sigma}(t)$ match the $r$ leading singular values of the flow field. To align the bases with the singular value decomposition representation, we rotate the in-plane and spanwise modes as
\begin{align}
\tilde{\mathsfbi{\Phi}} &= \mathsfbi{\Phi}\mathsfbi{U}, \\
\tilde{\mathsfbi{\Psi}} &= \mathsfbi{\Psi}\mathsfbi{Y}.
\end{align}
 Therefore, the evolved modes using Equations \ref{Phi_TDB} and \ref{Psi_TDB} are ranked based on energy. In our modal analysis, we visualize the rotated modes.} Examples of the evolving in-plane,  spanwise modes, and singular values are shown in figures~\ref{fig:TDB}$(c)$, $(d)$, and $(e)$, respectively. 
The leading in-plane modes represent the dominant flow structure, enabling interpretation of pattern changes over time. This study sets $r$ based on energy criterion \(1-\sum_{i=1}^{r}\sigma_i^{2}\big/\sum_{i}\sigma_i^{2}<10^{-4}\) at the first time step.

In summary, let us revisit the workflow of the present time-dependent bases analysis for the vortex–gust interaction using figure~\ref{fig:TDB}.
By matricizing the data, we initialize the time-dependent bases formulation using the singular value decomposition of the data.
We then advance the system through the time-integration to compute (i) the in-plane modes, (ii) the spanwise modes, and (iii) the matrix $\mathsfbi{T}(t)$ and its singular values to compare energy levels.
In practice, the in-plane modes reveal dominant structural features, such as shear-layer roll-up, vortex cores, and gust-induced separation/reattachment. 
Their energy content is quantified by the time-varying singular values of reduced covariance matrix, which track modal energy transfer. 
In particular, the modal energies correlate with fluctuations in the lift coefficient $C_L(t)$. 
Moreover, a large separation between the leading singular value and the rest indicates a more coherent flow during the gust–vortex interaction.

\section{Results}
\label{sec:results}

Let us apply the discussed modal analysis to extreme-vortex gust–airfoil interactions at a chord-based Reynolds number of $Re= u_\infty c/\nu =   5000$ \cite{fukami2025extreme}. Here, $c$ is the chord length and $\nu$ is the kinematic viscosity.
The datasets are produced with large-eddy simulations of a NACA0012 airfoil at an angle of attack \(\alpha=14^\circ\), yielding a turbulent separated wake.
The domain for the present data-driven analysis is defined in \((x,y,z) \in [-1.5,4.5]\times[-1.5,1.5]\times[0,1]\), with the leading edge at the origin. The grid size is \((N,M,K)=(300,150,50)\) and the time step is \(\Delta t=0.005\).
This study considers the convective time range of \(t \in [-2,\,3.5]\) covering the primary interaction process, with \(t=0\) marking the vortex–airfoil encounter.

To capture three-dimensional interaction dynamics, we use a spanwise width of \(1c\) with periodic boundaries in \(z\)~\cite{rolandi2024invitation,rolandi2025biglobal}.
An extremely strong disturbance is imposed with a Taylor-vortex gust at \(x_0/c=-2\)~\cite{taylor1918dissipation} as,
\begin{equation}
u_\theta=u_{\theta, \max } \frac{r_v}{R} \exp \left[\frac{1}{2}\left(1-\frac{r^2_{v}}{R^2}\right)\right].
\end{equation}
Where $r_v$ is the radial distance from the vortex center and $R$ is the vortex core radius, at which the vortex velocity $u_{\theta}$ reaches its maximum value ($u_{\theta,\max}$). In this paper, the disturbance strength and size are characterized by the gust ratio
$G \equiv u_{\theta,\max}/u_\infty$ and $D \equiv 2R/c$, respectively, where $u_\infty$ is the free-stream velocity.
Further details on numerical setups are referred to Fukami et al.~\cite{fukami2025extreme}.

\begin{figure}[t]
   \centerline{\includegraphics[width=1\textwidth]{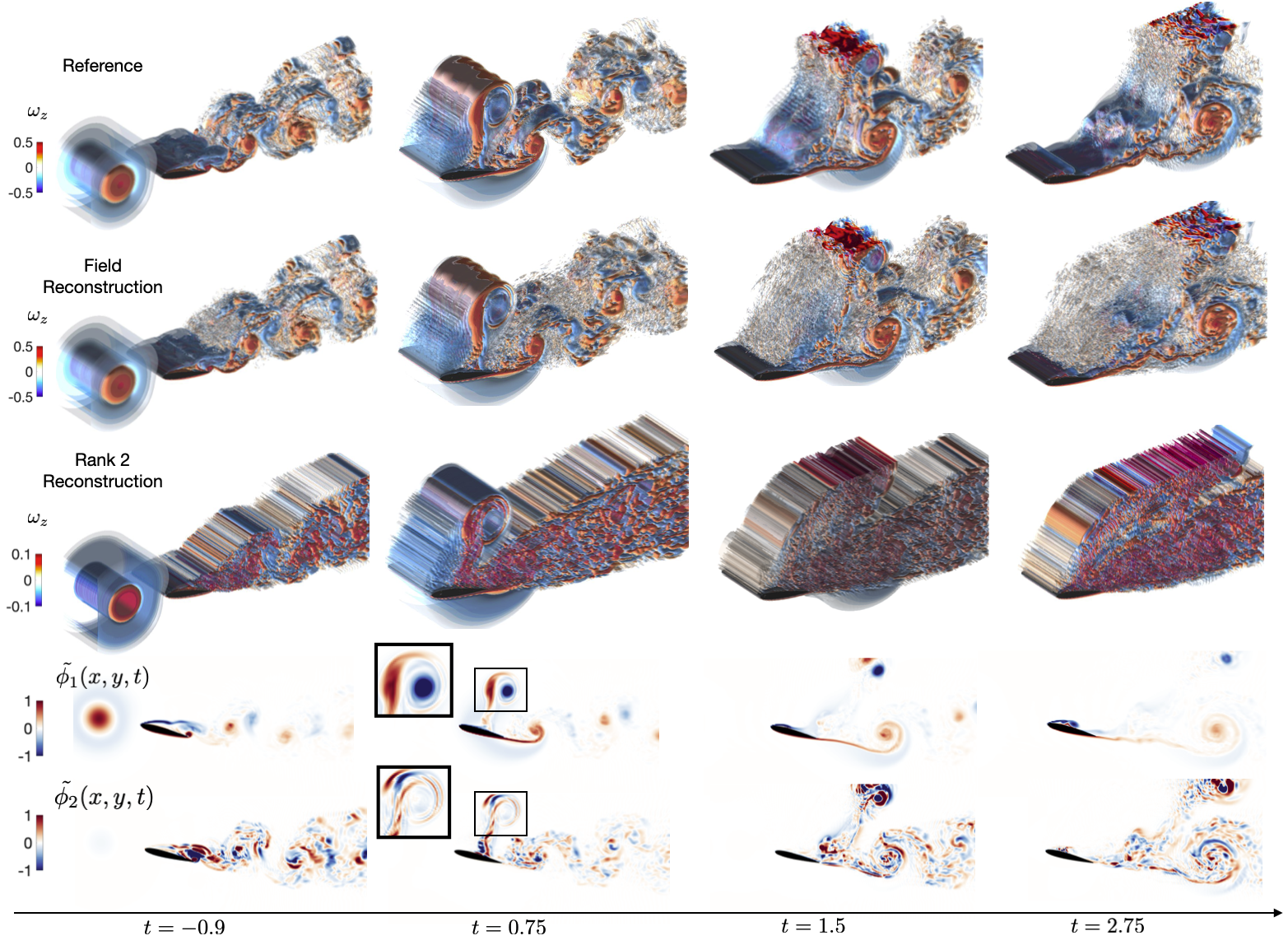}}
  \caption{Gust–airfoil interaction $(G,D)=(2,0.5)$. 
Reference spanwise vorticity field~\(\omega_z\), time-dependent bases reconstruction, rank-2 reconstruction, and in-plane modes are showed, respectively. }
\label{fig:G2}
\end{figure}

\begin{figure}[t]
   \centerline{\includegraphics[width=1\textwidth]{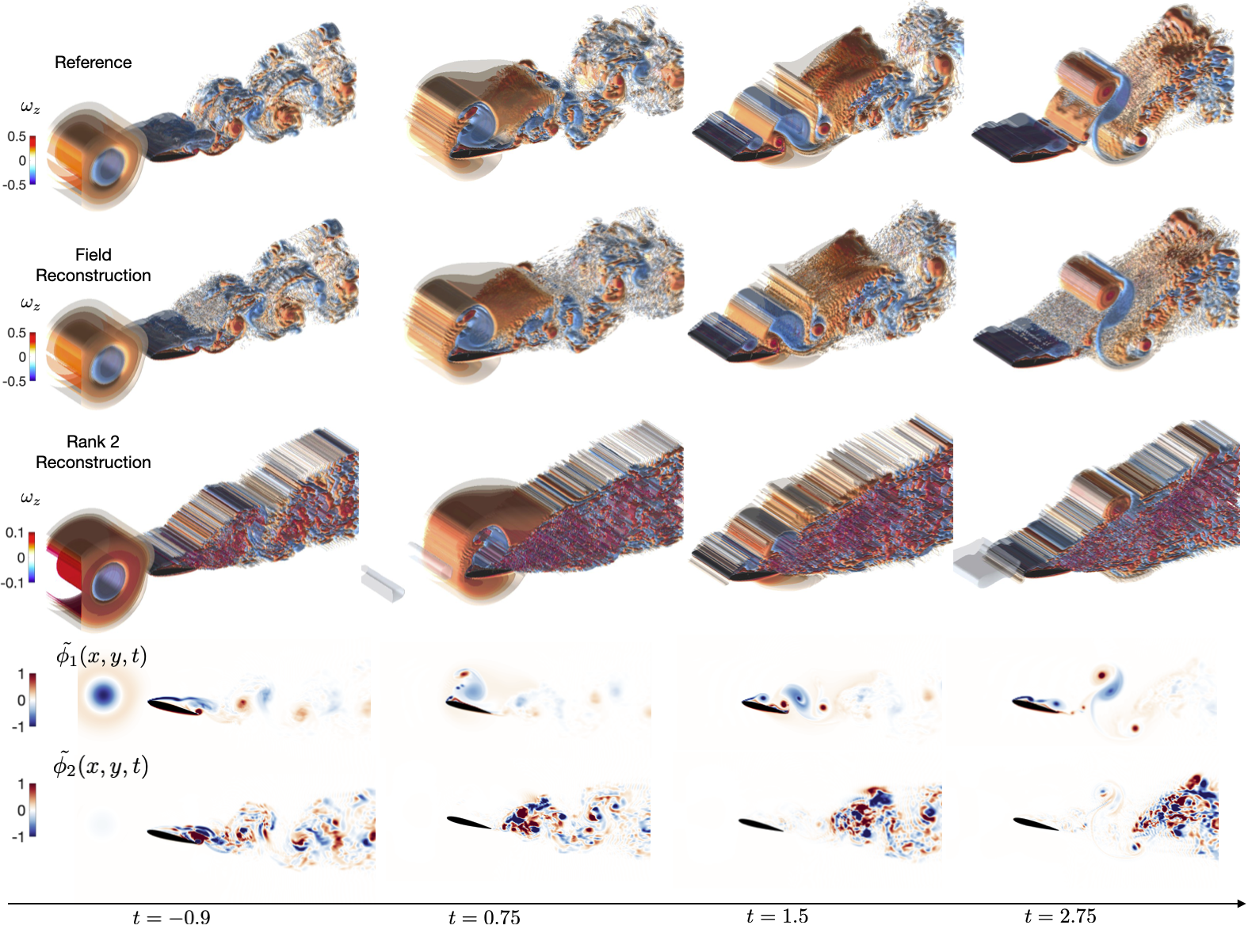}}
   \caption{Gust–airfoil interaction \( (G,D)=(-2,0.5)\). 
Reference spanwise vorticity field \(\omega_z\), time-dependent bases reconstruction, rank-2 reconstruction, and in-plane modes are showed, respectively.}
\label{fig:Gm2}
\end{figure}

We examine four cases $(G,D)\in\{(2,0.5),\,(-2,0.5),\,(2,1.5),\,(4,0.5)\}$, a range typically avoided in flight due to severe unsteadiness~\cite{fukami2023grasping}. In all cases, the vertical offset is \( Y \equiv y_0/c = 0.1 \).
This selection enables a comprehensive analysis by considering \( (G,D) = (2,0.5) \) as the baseline and study the effects of a negative gust \( (G,D) = (-2,0.5) \), a larger gust \( (G,D) = (2,1.5) \), and a more intense gust \( (G,D) = (4,0.5) \). 
We first focus on the baseline case with \((G,D)=(2,0.5)\), as presented in figure~\ref{fig:G2}.
An incoming vortex impinges on the leading edge, initiating shear-layer roll-up into a leading-edge vortex and briefly thickening separation.
The leading-edge vortex then pinches off and convects downstream, leaving a renewed shear layer and an organized wake of alternating vortical packets.
The reconstructed fields with the data-driven time-dependent bases are also shown under the series of the reference.
They preserve the main flow features over time, indicating that the present time-varying modes successfully capture the transient dynamics of extreme vortex-gust airfoil interactions.
The noisy behavior observed near the surface and within shear-layer regions at $t=2.75$  arises from accumulated time-integration and truncation errors with rank \(r=37\).

To further discuss what transient features are regarded as dominant, let us exhibit in figure~\ref{fig:G2} the rank-2 reconstruction with the two leading modes.
While capturing the dominant convectional movement of vortex cores over time, overall structures are smoothed out in the spanwise direction. 
In other words, the current data-driven approach assesses such two-dimensional physics as dominant over the present transient dynamics without the knowledge of the spatial length scale. 
This is similar to findings in previous studies based on nonlinear machine learning \cite{fukami2025extreme,fukami2025information}.

The current time-varying interaction dynamics between an extreme vortex gust and turbulent airfoil wakes can also be examined with the evolution of the first two in-plane modes, {\(\tilde{\phi}_{1}\) and \(\tilde{\phi}_{2}\)}.
Before impingement, the first mode {\(\tilde{\phi}_{1}\)} dominates, capturing the incoming vortex and the attached shear layer.
After impact, the second mode {\(\tilde{\phi}_{2}\)} gains energy near the leading edge and within the shear layer, consistent with the formation of the leading-edge vortex, temporary separation thickening, and the subsequent organization of wake packets.
The current observation of time-varying energy transfer across the length scale also coincides with the findings of a scale-decomposition approach~\cite{fujino2023hierarchy,fukami2025extreme}{, which captures the shear layer as the energy sender and receiver across the spatial length scales}.

We note that in-plane modes obtained through the current technique describe the time-varying dominant structures based on the energetic contents of transient flows.
In other words, there are cases where various structures across the spatial length scale appear in a single mode when they have a similar energy contribution, which happens in the present case with turbulent fine length scales.
One mode can reflect multiple dominant structures, such as shear layers, wake shedding, vortex interactions, and fine-scale turbulence, rather than isolating one structure per mode, since these events become energetically comparable.
As a result, the leading modes provide a compact summary of the primary energetic structure of unsteady flows.

\begin{figure}[t!]
\centerline{\includegraphics[width=1\textwidth]{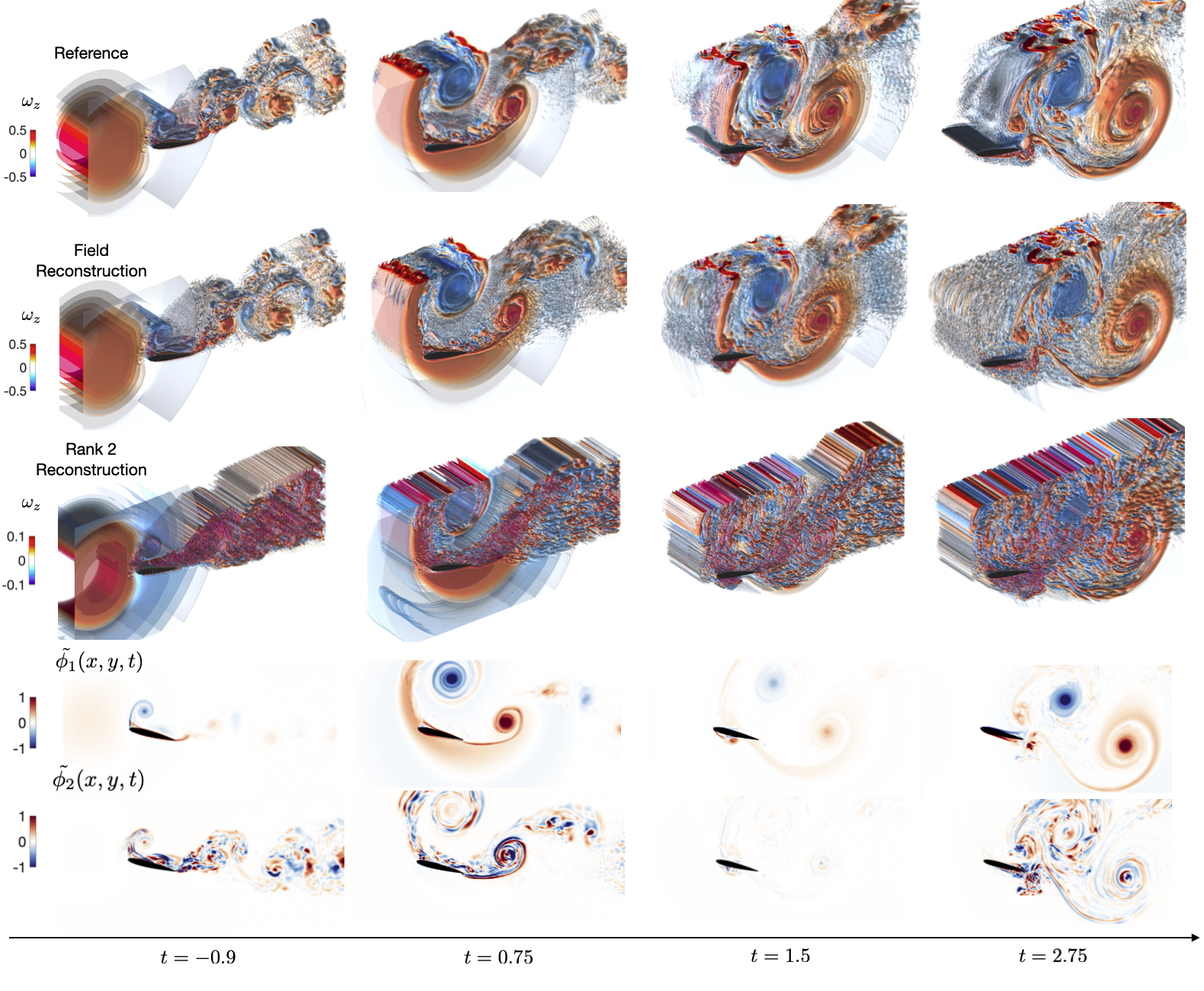}}
 \caption{Gust--airfoil interaction $(G,D)=(2,1.5)$.
          Reference spanwise vorticity field~\(\omega_z\), time-dependent reconstruction, rank-2 reconstruction, and in-plane modes are showed, respectively.}
\label{fig:G2D15}
\end{figure}

Let us examine the effect of the gust sign with \((G,D)=(-2,0.5)\), as shown in figure~\ref{fig:Gm2}. 
Relative to the case with a counter-clockwise vortex gust, the negative-gust impingement exhibits weaker undulations and a smoother streamwise organization~\cite{zhong2025optimally}.
The counter-rotating vortex thins and temporarily stabilizes the suction-side shear layer, delaying leading-edge vortex formation. 
These features are preserved by the reconstructed field shown under the reference field.
The noise observed at \(t=2.75\), near the surface and within the shear layer, is reduced relative to the positive gust case, indicating slower error growth. 
This occurs because the stabilized shear layer and delayed leading-edge vortex formation generate less small-scale content and weaker gradients, enabling the retained modes (\(r=37\)) to capture the dynamics more accurately.
The dominant convection of vortex cores is captured by the two-leading-mode reconstruction, highlighting the transient features over time where the spanwise variations are smoothed, consistent with two-dimensional vortical structure. 
A faint lobe that emerges ahead of the leading-edge vortex in the rank-2 reconstruction at $t=0.75$ and 2.75 is likely caused by low-rank truncation, which disappears as rank increases.

The interaction between the extreme vortex gust and the turbulent airfoil wake is evident in the evolution of the two leading in-plane modes. 
Before the vortex impingement, the first mode {\(\tilde{\phi}_{1}\)} concentrates on the vortex core and the attached shear layer, whereas the second mode {\(\tilde{\phi}_{2}\)} carries little energy. 
After impingement, {\(\tilde{\phi}_{1}\)} tracks the convective motion of the leading edge vortex, while {\(\tilde{\phi}_{2}\)} shows no clear signature of the shear layer.
Compared with the positive-gust case, {\(\tilde{\phi}_{2}\)} remains weaker because of less small-scale content. 
A persistently weak {\(\tilde{\phi}_{2}\)} indicates a lower dimensional state, with energy concentrated in {\(\tilde{\phi}_{1}\)} and reduced small-scale activity.
{Therefore, the dominant flow structures are well represented by the first leading mode in this case.}

\begin{figure}[t!]
   \centerline{\includegraphics[width=1\textwidth]{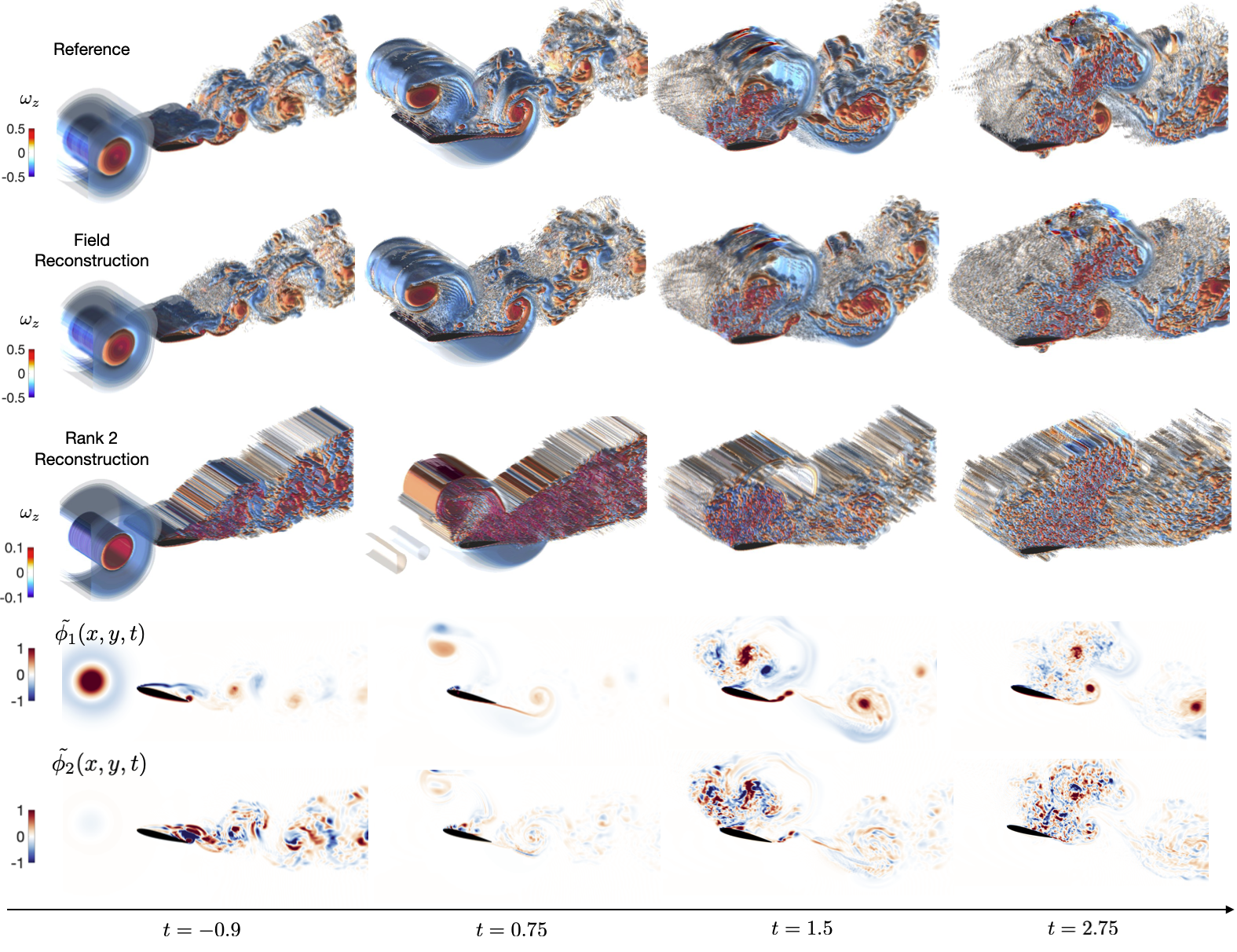}}
  \caption{Gust--airfoil interaction \((G,D)=(4,0.5)\).
Reference spanwise vorticity field~\(\omega_z\), time-dependent bases reconstruction, rank-2 reconstruction, and in-plane modes are showed, respectively.}
\label{fig:G4}
\end{figure}

Motivated by the extended interaction length and longer-lived separation of a larger gust, we also examine a larger gust of \((G,D)=(2,1.5)\), as depicted in figure~\ref{fig:G2D15}. 
Increasing the gust size advances the onset of interaction and produces longer-lived separated structures.
The time-dependent bases reconstruction with rank \(r=34\) preserves the main features but exhibits stronger noise and faster error growth over time relative to the baseline. 
This likely arises from steeper gradients around the leading-edge vortex and in the near wake. 
The added small-scale activity leads to increased energy in the second mode, reducing what the retained bases capture and accelerating truncation-error growth. 
The rank-2 reconstruction, recovering the large-scale placement and convection of the vortex core while smoothing spanwise structure, indicates that two-dimensional vortical structures are regarded as transient dominant feature in this larger-gust case as well.

To reveal the structures governing the interaction of a larger gust, we show the evolution of the two leading in-plane modes. Because the larger gust vortex advances the onset of interaction, an upstream vortex is only faintly visible. The first mode {\(\tilde{\phi}_{1}\)} emphasizes the leading-edge region and then carries the growing, convecting leading-edge vortex. The second mode {\(\tilde{\phi}_{2}\)} remains secondary, emphasizing shear-layer ripples and local wake adjustments near the leading edge vortex. 
Relative to the case with $D=0.5$, both modes capture the vortex core and shear layer, and  {\(\tilde{\phi}_{2}\)} gains more energy after impingement due to more violent interactions. 
{The comparison between the reference field and the in-plane modes at $t=2.75$ reveals that the baseline case exhibits weaker large-scale roll-up, whereas the larger-gust case produces a stronger rolled-up wake vortex. 
This demonstrates that the in-plane modes providing compact bases of the present transient vortex-airfoil interactions enable the analysis across not only time but also a range of gust cases, by highlighting where the dominant energetic contents concentrate and how they reorganize across the gust parameters.}

The effect of gust ratio is also analyzed with (\(G,D)=(4,0.5)\), as presented in figure~\ref{fig:G4}.
The leading-edge vortex wraps over the suction side and re-impinges on the airfoil, and the wake becomes less organized, compared to the case with $G=2$, where separation is briefer, the leading-edge vortex convects downstream after a single pass, and the wake remains organized. 
As shown under the reference, the time-dependent bases reconstruction with \(r=37\) preserves the large-scale evolution, but exhibits faster error growth than the baseline, corresponding to the richer small-scale content and steeper gradients. 
Despite the complexity of the flow with turbulent structures, the rank-2 reconstruction captures the overall trends of the convection and smooths the spanwise direction, capturing the two-dimensional physics of the problem.

Examining the two leading in-plane modes clarifies the dynamics in the stronger-gust case. {\(\tilde{\phi}_{1}\)} dominates before impingement. {\(\tilde{\phi}_{2}\)} engages earlier and more strongly than for \(G=2\) following the impact, capturing leading-edge vortex roll-up, the re-impingement trajectory, and an intensified, longer-lived separated region. Consequently, both modes persist longer than in the weaker gust core, consistent with the stronger, multi-stage interaction at \(G=4\). 
{Comparing the reference spanwise vorticity field across cases at $t=2.75$, the strongest-gust case exhibits increased fine-scale activity in the wake and reduced coherence.
The in-plane modes at the same time reflect this increment of spatiotemporal complexities as mixed, less distinctly organized structures in the wake region.}

\begin{figure}[t!]
    \includegraphics[width=0.95\textwidth]{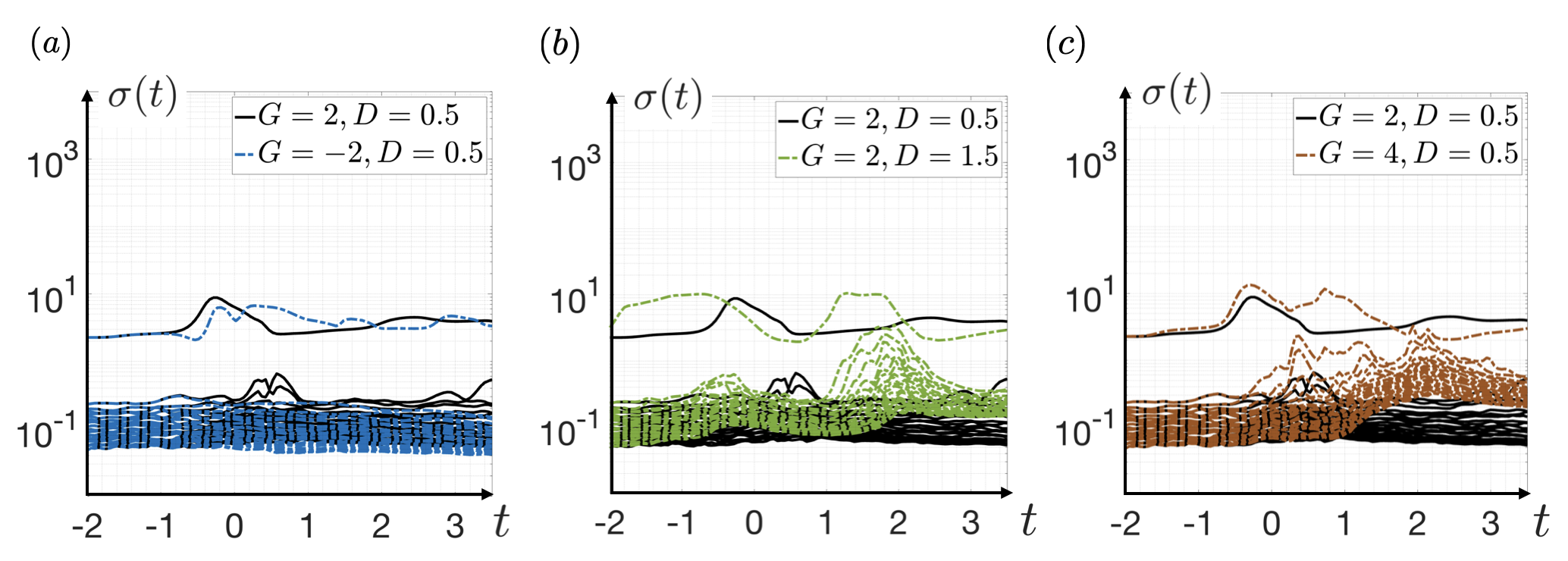}
\caption{{Singular value evolution comparison for all cases.}}
\label{fig:SVC}
\end{figure}
\begin{figure}[t!]
\centering
    \includegraphics[width=0.8\textwidth]{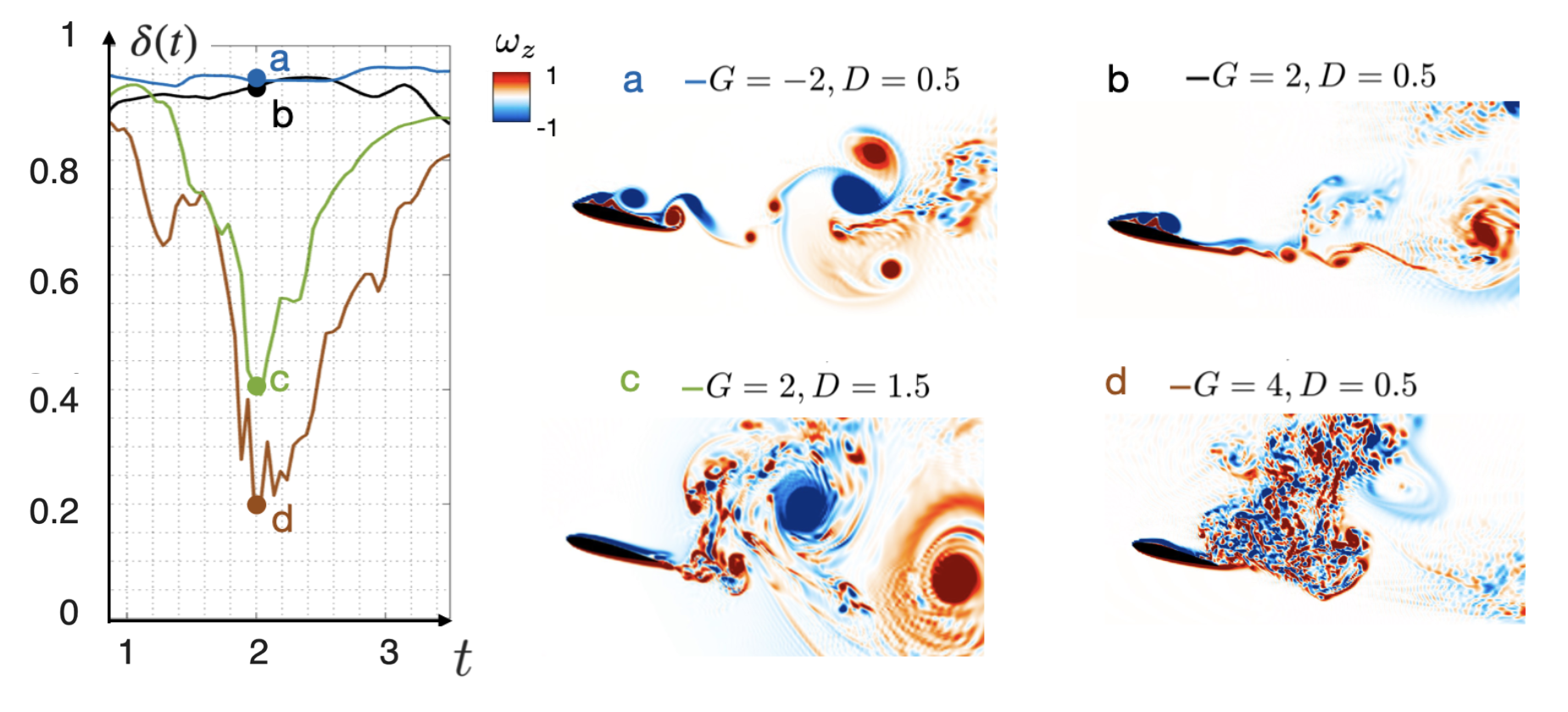}
\caption{
        {Quantifying coherent structure based on the relative singular value gap.}}
\label{fig:SVC_Dist}
\end{figure}

\begin{figure}[t!]
    \centering
    \includegraphics[width=0.95\textwidth]{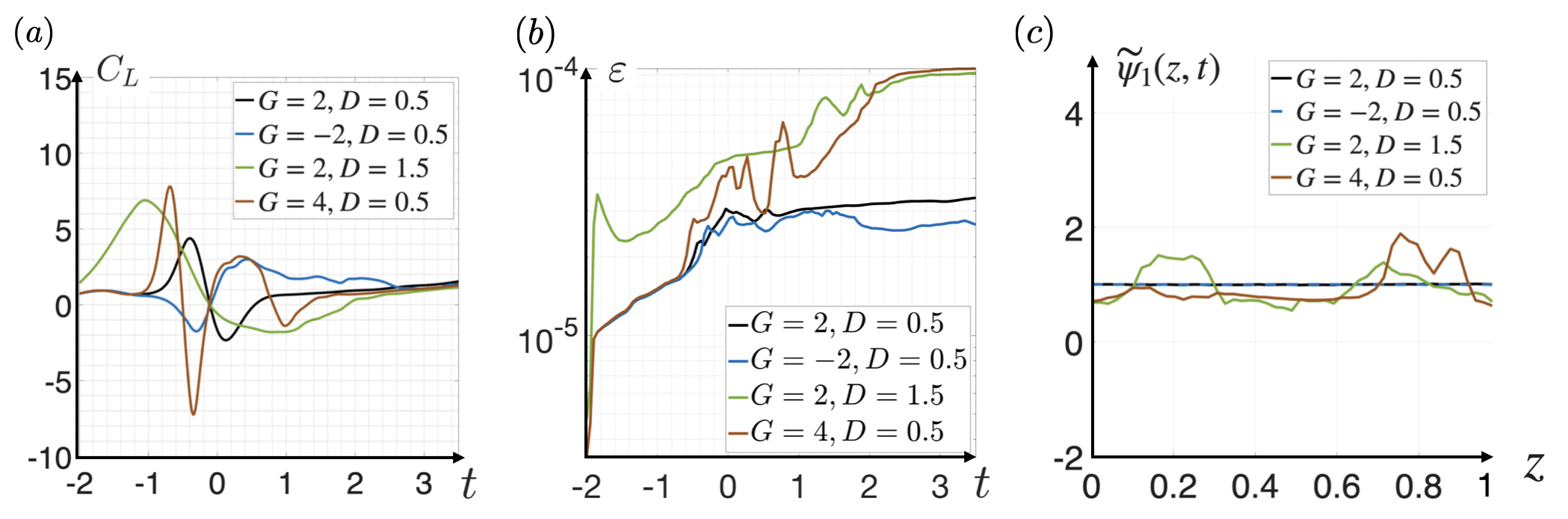}
    \caption{{  Comparison of $(a)$ lift coefficient \(C_L\), $(b)$ reconstruction error, and $(c)$ the leading spanwise modes ($t=2$).}}
    \label{fig:Cl-er}
\end{figure}

{In addition to the structures visualized in the in-plane modes, structural coherence of the present transient turbulent flows can also be quantified by comparing the modal energy based on singular values.}
As shown in figures~\ref{fig:SVC}$(a-c)$, the smaller singular values \((\sigma_2,\ldots,\sigma_r)\) gain energy for the stronger and larger gusts after impingement and nearly approach the leading singular value (\(\sigma_1\)).
This rise reduces the gap to the leading singular value, which represents the mean/large-scale component, so that, as the smaller values approach it, they become comparably important. Toward the end of the time domain of interest, all singular values decay, indicating wake reorganization. 
At the same time, the decreased separation between the leading singular value and the others reflects reduced coherence and stronger multiscale activity. 
These trends correspond to the coherent-structure patterns in the final snapshot of figures~\ref{fig:G2}–\ref{fig:G4}, where \(G=-2\) appears most coherent and \(G=4\) less coherent, {due to reimpingement}. {The behavior of singular values and their connection with coherent structure can be 
quantified by the relative singular value gap $\delta(t) = (\sigma_1(t) - 
\sigma_2(t)) / \sigma_1(t)$, which measures the dominance of the leading singular values 
relative to the rest of the spectrum. When $\delta(t) \to 1$, the flow has more 
coherent structure, whereas $\delta(t) \to 0$ indicates that the smaller-scale 
structures are gaining energy equivalent to the leading singular value, reflecting a more 
complex and less organized flow. As shown in Figure~\ref{fig:SVC_Dist}, the negative gust case ($G = -2$) and the 
baseline maintain a large relative gap throughout the time interval, indicating that 
the flow retains a dominant coherent structure even after the gust impact. In 
contrast, increasing the gust strength leads to a significant drop in $\delta(t)$, 
reflecting the multi-scale vortical 
interactions. The most intense gust case ($G = 4$) exhibits the smallest relative 
gap, corresponding to the most disorganized and least coherent flow due to reimpingement. 
This is consistent with the vorticity field observed in the flow 
snapshots.}

The present data-driven technique captures the aerodynamic features of the flow from vorticity data, as shown in figure~\ref{fig:Cl-er}$(a)$.
The peaks of the leading singular value coincide with the lift coefficient.
For \(G=2\) and \(G=-2\), the lift responses have similar magnitudes but are mirrored by the gust sign; in the negative case, the leading singular value remains slightly larger after the encounter (more retained energy), thus \(C_L(t)\) decays more slowly.
Increasing the gust size to \(D=1.5\) amplifies the peak–trough excursion and extends the duration of post-impingement oscillations in lift, consistent with the trend of the first singular value and a rise in the smaller singular values. 
For the strongest-intensity case \((G=4)\), the smaller singular values approach the leading singular value and remain elevated (the spectrum does not fully relax), producing the largest overshoot/undershoot and long-lived, multi-frequency lift oscillations. The rise in the singular values also indicates a fast growth rate of the error, as shown in figure~\ref{fig:Cl-er}$(b)$ in which the error is defined as
\(
\varepsilon = \left\lVert \mathsfbi{V} - \hat{\mathsfbi{V}} \right\rVert_{2}/ (N \times M \times K)
\). 
Under stronger multiscale activity, smaller modes gain more energy but are truncated, hence the unresolved content grows.

\begin{figure}[t!]
    \centering
    \includegraphics[width=0.85\textwidth]{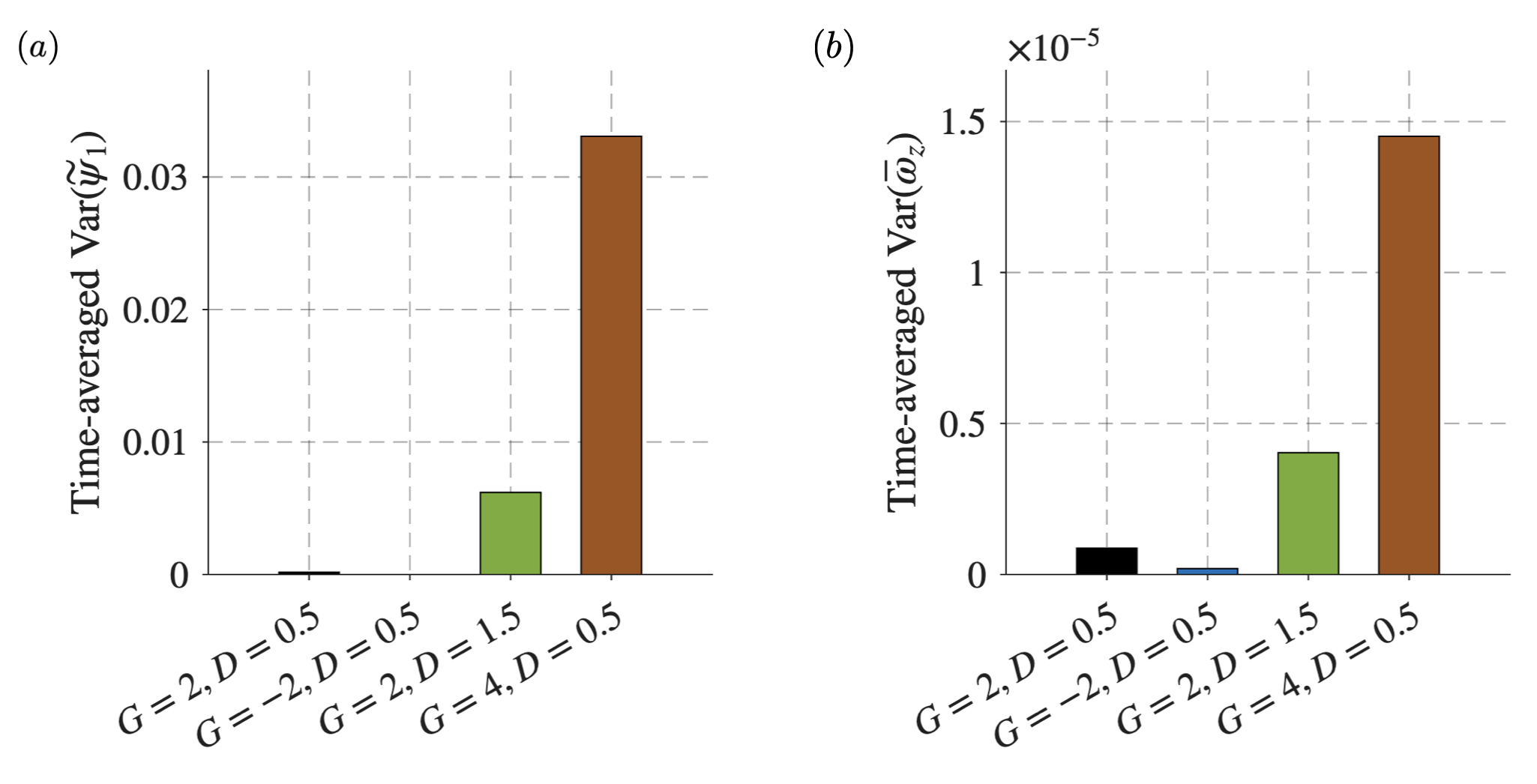}
    \caption{{Time-averaged spanwise variance of (\textit{a}) $\tilde{\psi}_1$ and (\textit{b}) $xy$-averaged vorticity $\bar{\omega}_z$.}}
    \label{fig:var}
\end{figure}

{The structure of gust impingement is described with in-plane modes. 
In-plane modes reveal the structure in the $xy$-plane, while spanwise 
modes show structural variation in $z$. For the negative and baseline gusts, 
$\tilde{\psi}_1(z,t)$ is nearly constant, implying that the first in-plane 
mode approximates the spanwise average of the field in a normalized manner. 
Given that the leading singular value is much larger than the others, the main 
features of the flow are nearly two-dimensional, as shown in 
figures~\ref{fig:G2}--\ref{fig:Gm2}. This also holds for the larger and stronger 
gust; however, at certain instants the relative gap between the singular values 
reduces, as shown in figure~\ref{fig:SVC_Dist}, and the first spanwise mode 
begins to change, as shown in figure~\ref{fig:Cl-er}$(c)$. 
Figure~\ref{fig:var} further quantifies this variation through the 
time-averaged spanwise variance of the leading spanwise mode 
$\mathrm{Var}(\tilde{\psi}_1)$, panel~(a), and the spanwise variance of the 
$xy$-averaged vorticity $\mathrm{Var}(\bar{\omega}_z)$, panel~(b). Both 
quantities increase with gust intensity, with $G=4$, $D=0.5$ showing the 
largest spanwise variation and $G=-2$, $D=0.5$ the smallest. Although the 
variations are small, confirming that the flow remains largely two-dimensional, 
the results indicate that more intense gusts produce increasingly 
three-dimensional behaviour.} These results suggest that the current technique enables data-driven transient modal analysis of the interaction between extreme gusts and turbulent airfoil wake, while providing time-varying spatial modes and singular values as a complexity indicator.

\section{Concluding remarks}
\label{sec:conclusion}

We examine extreme vortex–gust interactions over a NACA0012 airfoil at $Re=5000$ through data-driven time-dependent bases {to analyze the extreme aerodynamic conditions arising from wing--gust 
interaction.}  
The current technique establishes a link between the flow pattern and the evolution of in-plane modes{, spanwise modes,} and modal energy, {providing 
a compact framework for interpreting the complex flow dynamics.}
We focus on in-plane modes and their singular values because they capture the dominant structures with the highest energy content, yielding a compact and interpretable description of the {vortical} structure. 
The first mode captures dominant convection while the second mode gains energy after the vortex impingement, which is further amplified as the gust becomes stronger and larger.
In the present data-driven technique, aerodynamic flow features are obtained from vorticity data; in particular, the peaks of the leading singular value and the lift coefficient occur at the same time. 
Moreover, the singular value spectrum quantifies structural coherence: a large separation between the leading singular value (representing the mean/large-scale component) and the rest indicates high coherence, whereas a reduced gap signals diminished coherence and richer multiscale activity. {Furthermore, changes in the leading spanwise mode indicate 
the appearance of three-dimensional flow behaviour. When the flow is nearly 
two-dimensional, the leading spanwise mode remains approximately constant along $z$ 
and the singular values exhibit a significant gap; spanwise variation in 
the leading spanwise mode and a reduced singular value gap show the onset of 
three-dimensionality. This work focuses on modal analysis applicable to transient gust 
encounters exhibiting an unsteady base state, enabling the low-order quantification 
of complex vortex interactions. The performance of time-dependent 
bases decomposition can be further extended to fully three-dimensional gust 
encounters in the future~\cite{odaka2026vortical,odaka2025extreme}. The analysis of multiple sequential gust interactions also represents a direction for future framework~\cite{fukami2023grasping}. Future work may also explore the influence of varying gust frequency and profile on the modal decomposition.}
In addition, one can investigate sparse-sensor reconstruction \cite{naderi2023adaptive}, where time-dependent bases are inferred from sparse spatial measurements and the time dependence enables adaptive updating of the sampled points as the flow evolves.
The current technique may offer a data-driven foundation for transient modal analysis of a range of aerodynamic flows with unsteady base states.

\section*{Acknowledgments}
K.F. acknowledges support from the JSPS KAKENHI Grant No. JP25K23418 and No. JP26K01129, the JST PRESTO Grant No. JPMJPR25KA, and the MEXT Coordination Funds for Promoting Aerospace Utilization Grant No. JPJ000959.

\section*{Declaration of interests}

{The authors report no conflict of interest.}

\bibliographystyle{unsrt}  
\bibliography{references}

\end{document}